\documentclass[%
 reprint,
superscriptaddress,
 amsmath,amssymb,
 aps,
prapplied,
floatfix,
]{revtex4-2}

\usepackage{graphicx}
\usepackage{dcolumn}
\usepackage{bm}
\usepackage{bbm}
\usepackage{textcomp}
\usepackage[utf8]{inputenc}
\usepackage[colorlinks]{hyperref}

\begin{document}

\preprint{APS/123-QED}

\title{Eigenmode and field imaging of a compact superconducting resonator array}

\author{Benedek Ga\'al}
 \email[Contact author: ]{benedek.gaal@epfl.ch}
 \affiliation{Institute of Physics, École Polytechnique Fédérale de Lausanne (EPFL), CH-1015 Lausanne, Switzerland}
 \affiliation{Center for Quantum Science and Engineering, École Polytechnique Fédérale de Lausanne (EPFL), CH-1015 Lausanne, Switzerland}
 
\author{Vincent Jouanny}
 \affiliation{Institute of Physics, École Polytechnique Fédérale de Lausanne (EPFL), CH-1015 Lausanne, Switzerland}
 \affiliation{Center for Quantum Science and Engineering, École Polytechnique Fédérale de Lausanne (EPFL), CH-1015 Lausanne, Switzerland}

\author{Simone Frasca}
 \altaffiliation[Also at: ]{Department of Physics, ETH Zurich, CH-8093 Zurich, Switzerland}
 \affiliation{Institute of Physics, École Polytechnique Fédérale de Lausanne (EPFL), CH-1015 Lausanne, Switzerland}
 \affiliation{Center for Quantum Science and Engineering, École Polytechnique Fédérale de Lausanne (EPFL), CH-1015 Lausanne, Switzerland}
 
\author{Christophe Galland}
 \affiliation{Institute of Physics, École Polytechnique Fédérale de Lausanne (EPFL), CH-1015 Lausanne, Switzerland}
 \affiliation{Center for Quantum Science and Engineering, École Polytechnique Fédérale de Lausanne (EPFL), CH-1015 Lausanne, Switzerland}

\author{Pasquale Scarlino}
 \affiliation{Institute of Physics, École Polytechnique Fédérale de Lausanne (EPFL), CH-1015 Lausanne, Switzerland}
 \affiliation{Center for Quantum Science and Engineering, École Polytechnique Fédérale de Lausanne (EPFL), CH-1015 Lausanne, Switzerland}
 
\author{Valentin Goblot}
 \affiliation{Institute of Physics, École Polytechnique Fédérale de Lausanne (EPFL), CH-1015 Lausanne, Switzerland}
 \affiliation{Center for Quantum Science and Engineering, École Polytechnique Fédérale de Lausanne (EPFL), CH-1015 Lausanne, Switzerland}

\date{\today}

\begin{abstract}
Superconducting resonator arrays can tailor the photonic environment for quantum information processing, but conventional port-based microwave spectroscopy alone does not resolve the rich spatial structure of their modes. Here, we apply laser scanning microscopy (LSM) and nitrogen-vacancy (NV) center magnetometry to a two-dimensional superconducting resonator array consisting of 33 lumped-element resonators arranged in a honeycomb lattice hosting topological edge states. We demonstrate the complementarity of the two techniques: LSM images the eigenmodes, while NV magnetometry maps the local microwave magnetic field amplitude. Using this spatial information, we experimentally validate that the realized lattice matches the design, we image the topological edge state and measure its sublattice polarization, and place an upper bound of $\approx0.3\%$ on the relative on-site frequency disorder in the lattice. Our work establishes a combined spatial characterization approach for compact superconducting resonator arrays with applications in hybrid spin--superconducting quantum systems.
\end{abstract}

\maketitle


\begin{figure*}[t]
    \includegraphics[width=0.95\linewidth]{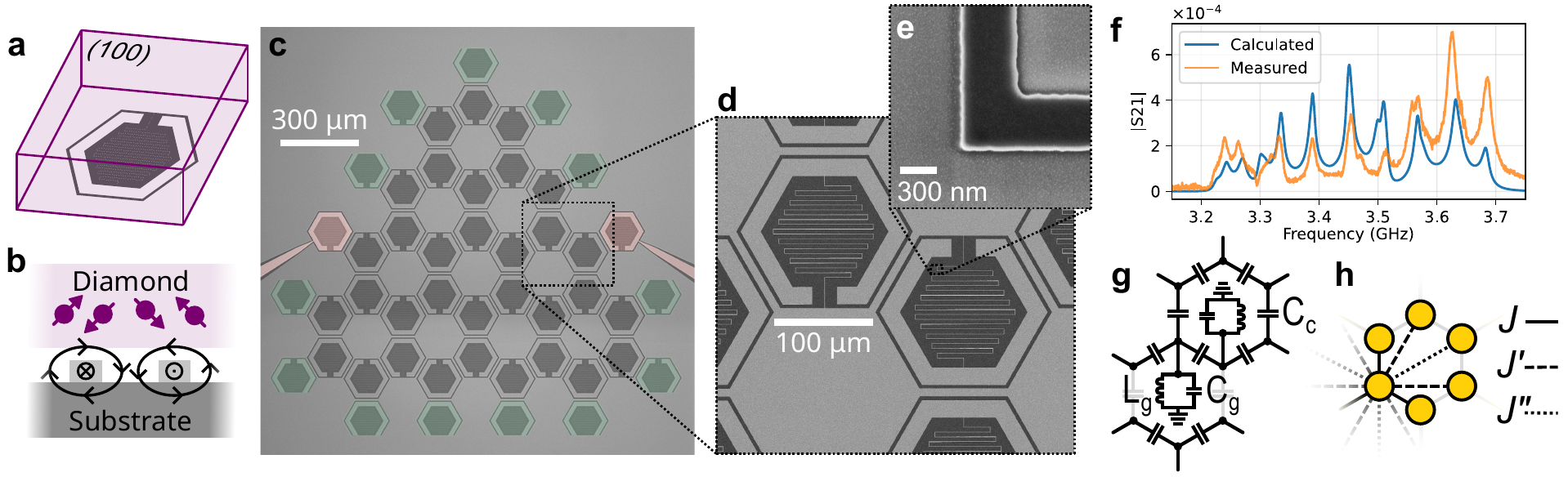}
    \caption{\textbf{Two-dimensional CCA.} \textbf{a} Sketch of the experimental configuration: we integrate spins with superconducting resonators by placing a slab of diamond on top. \textbf{b} A thin layer of NV centers near the diamond surface couple to the near-field RF magnetic field of the inductors. \textbf{c} Optical micrograph of the resonator array. The CCA consists of 33 identical lumped-element resonators arranged in a honeycomb pattern, with ``ghost resonators'' at the edge (highlighted in green) and two coupling ports (highlighted in red). \textbf{d-e} Scanning electron micrograph of the resonators and the inductor. \textbf{f} Measured (orange) and simulated (blue) RF spectrum of the device. \textbf{g} Schematic of the lumped-element model of the CCA. Each cavity is modelled as an LC resonator with capacitance to ground $C_g$, inductance to ground $L_g$, coupling capacitances $C_c$, and total capacitance $C_{\Sigma}=C_g+\sum C_c$. The parasitic next-nearest neighbour capacitances are not shown. \textbf{h} Representation of the corresponding tight-binding Hamiltonian, with nearest ($J$), next-nearest ($J'$) and next-next-nearest ($J''$) neighbour couplings.}
    \label{fig:CCA}
\end{figure*}

\section{Introduction}\label{sec:intro}



Arranging superconducting microwave resonators into coupled arrays offers the possibility to engineer microwave photonic band structures on-chip, with great flexibility~\cite{carusottoPhotonicMaterialsCircuit2020}. Such coupled cavity arrays (CCAs) have enabled solid-state analog quantum simulators~\cite{houckOnchipQuantumSimulation2012, zhangSuperconductingQuantumSimulator2023} and quantum information processing technologies~\cite{naikRandomAccessQuantum2017, ferreiraDeterministicGenerationMultidimensional2024}. CCAs also provide a structured photonic bath for other quantum systems, forming the basis of hybrid platforms~\cite{clerkHybridQuantumSystems2020}; notable examples include integration with superconducting qubits~\cite{kimQuantumElectrodynamicsTopological2021, jouannySuperstrongDynamicsDirectional2025}, mechanical oscillators~\cite{youssefiTopologicalLatticesRealized2022}, or spin ensembles~\cite{kuboStrongCouplingSpin2010}.
 
Early superconducting microwave CCAs relied on conventional superconducting materials and were implemented using both coplanar-waveguide resonators~\cite{carusottoPhotonicMaterialsCircuit2020, underwoodImagingPhotonLattice2016, kollarHyperbolicLatticesCircuit2019} and lumped-element resonators~\cite{kimQuantumElectrodynamicsTopological2021,wangModeStructureSuperconducting2019}, enabling the realization of a wide variety of tight-binding Hamiltonians.
However, such resonators have a large physical footprint, which limits their scalability.
High-kinetic-inductance films such as NbN and NbTiN have emerged as a means to miniaturize resonators to the deep sub-wavelength, $100~\text{µm}$ scale, enabling compact lattices of lumped-element resonators that retain high quality factors~\cite{frascaHighkineticInductanceNbN2023}. 
In a previous work, we demonstrated successful implementation of one-dimensional CCAs made of such resonators~\cite{jouannyHighKineticInductance2025}. However, as lattices grow in complexity, radio-frequency (RF) spectroscopy alone becomes insufficient: fabrication disorder, defects, and unintended couplings reshape the eigenmodes and, with them, the coupling to external quantum systems. These effects are difficult to disentangle based on the transmission or reflection spectrum, calling for spatially resolved characterization.

Laser scanning microscopy (LSM) provides a perturbative method to obtain spatially resolved eigenmode amplitudes of resonator arrays: a focused laser beam locally raises the surface resistance through heating and quasiparticle generation \cite{zhuravelUnconventionalRfPhotoresponse2012,budoyoEffectsNonequilibriumQuasiparticles2016}, and the induced change in the RF spectrum reports the local mode weight \cite{wangModeStructureSuperconducting2019, morvanObservationTopologicalValley2021, morvanBulkPropertiesHoneycomb2022}. LSM has been applied to spatially characterize superconducting resonators and metamaterials \cite{zhuravelMeasurementLocalReactive2006, zhuravelUnconventionalRfPhotoresponse2012, wangModeStructureSuperconducting2019}, and to reconstruct the band structure and topological edge states of honeycomb lattices \cite{morvanObservationTopologicalValley2021,morvanBulkPropertiesHoneycomb2022}. 
Scanning defect microscopy is another perturbative approach that has demonstrated similar capabilities~\cite{underwoodImagingPhotonLattice2016}. 
The main limitation of such perturbative approaches is that the probe modifies the device and that it only provides indirect information on the local field. 

The negatively charged nitrogen-vacancy (NV$^-$) center in diamond offers a complementary, non-perturbative optically addressable probe of microwave magnetic fields. The spin-dependent quantum yield of fluorescence excited from the NV$^-$ spin-triplet ground state allows for optical detection of magnetic fields, from DC to the GHz range~\cite{barrySensitivityOptimizationNVdiamond2020}. The simplest AC magnetometry protocol relies on tuning the $m_s=0\longrightarrow \pm 1$ spin transitions in resonance with the target microwave signal, and extracting the resulting spin Rabi oscillation frequency. Spatially-resolved imaging of the local microwave field amplitude and direction can be obtained by nanoscale imaging with a single scanned spin \cite{appelNanoscaleMicrowaveImaging2015, wangHighresolutionVectorMicrowave2015} or wide-field imaging with near-surface ensembles \cite{shaoWideFieldOpticalMicroscopy2016,horsleyMicrowaveDeviceCharacterization2018,bassoWidefieldMicrowaveMagnetic2025}. This method captures the device's driven response under unperturbed operating conditions, making it well suited to driven-dissipative systems. 

In this work, we apply both techniques to image the eigenmodes of a two-dimensional array of deep sub-wavelength superconducting resonators. We realize an array of 33 lumped-element LC resonators from high-kinetic-inductance NbTiN, arranged in a honeycomb lattice with zigzag edges (Fig.~\ref{fig:CCA}a--c). Such a lattice is known to exhibit edge modes localized on the zigzag boundary~\cite{castronetoElectronicPropertiesGraphene2009, fujitaPeculiarLocalizedState1996, nakadaEdgeStateGraphene1996}, which have been proposed as a resource for cavity quantum electrodynamics with emitters coupled to the lattice edge~\cite{dibenedettoEmergentCavityQEDDynamics2026}. We first use LSM to image the field profiles of all port-coupled eigenmodes, revealing an edge-type mode polarized on the sublattice forming the zigzag boundary. From the same dataset, we extract an upper bound of $0.3\%$ on the on-site frequency disorder. We then interface the CCA with a diamond hosting NV centers (Fig.~\ref{fig:CCA}g,h) and perform AC magnetometry imaging of the microwave magnetic field amplitude under several driving conditions. 
Our results validate the controlled emulation of 2D Hamiltonians with compact superconducting resonator arrays and demonstrate a combined characterization approach, pivotal for scaling such arrays to larger resonator numbers and more complex geometries.

\begin{figure*}[t]
    \includegraphics[width=0.98\linewidth]{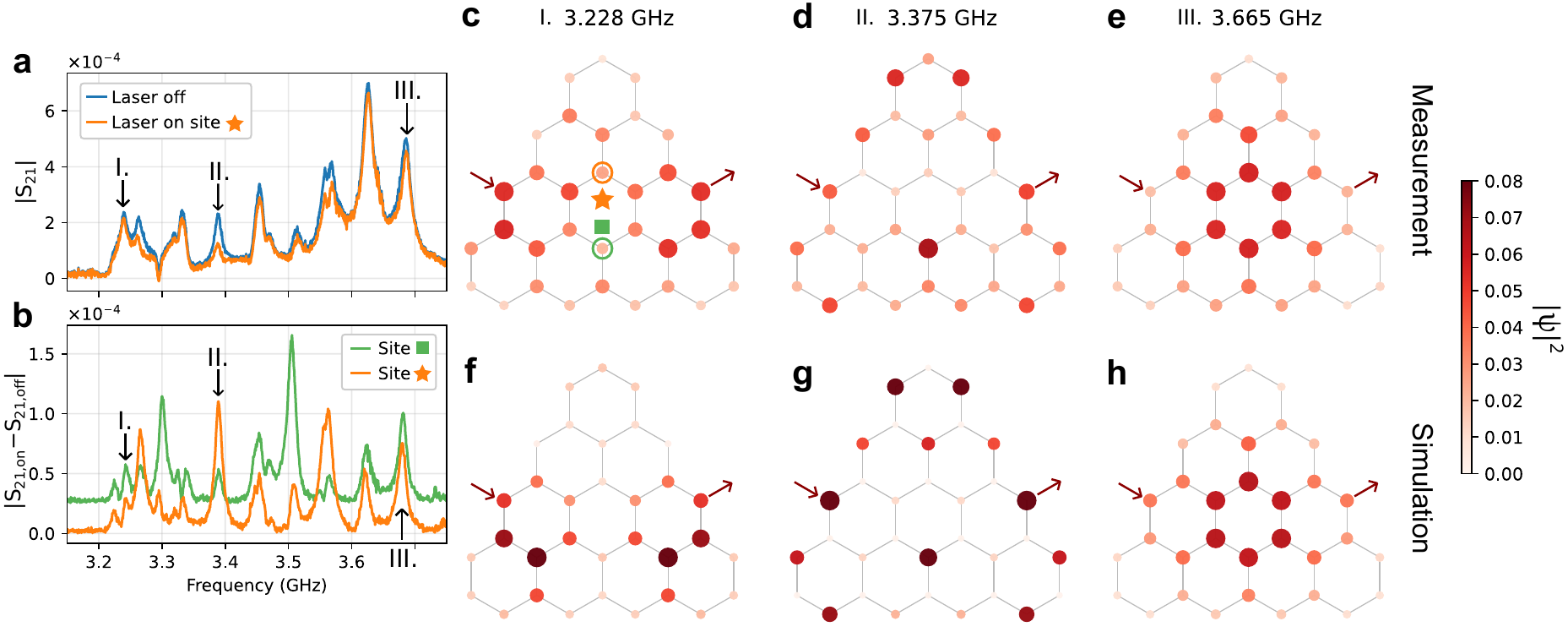}
    \caption{\textbf{LSM measurement of the device without diamond.} \textbf{a} RF spectrum of the device without (blue curve), and with (orange) $1.2~\text{µW}$ laser illumination of the specified site. \textbf{b} LSM differential signal for two specified sites. For each site, the peaks in the differential signal are identified and integrated, allowing us to construct a spatial map. \textbf{c--e} LSM spatial maps of three modes. The corresponding mode frequencies are highlighted with arrows in panels \textbf{a,b}, and the input and output ports are indicated by the arrows in the spatial maps. \textbf{f--h} Simulated LSM profiles of the same modes. }
    \label{fig:LSM1}
\end{figure*}

\section{Two-Dimensional Coupled Cavity Arrays}\label{sec:cca}

We design the resonator array to couple to an ensemble of NV centers (Fig.~\ref{fig:CCA}a, b), with its resonance frequency matched to the NV ground-state spin transition at $\approx2.87~\mathrm{GHz}$. We use NbTiN thin films with $\mathrm{T}_c$ around $12~\mathrm{K}$ and kinetic inductance of $7~\mathrm{pH/\square}$. We design hexagonal base units to make tileable resonators (Fig.~\ref{fig:CCA}c, d), made of two main components: a capacitive strip which contributes both to capacitance to ground and to coupling capacitances, and a thin inductive component which contributes to inductance to ground (Fig.~\ref{fig:CCA}e). Compared to our previous work on high kinetic inductance cavity arrays \cite{jouannyHighKineticInductance2025} we use larger capacitive pads to offset the lower kinetic inductance of NbTiN, and we also implement ground connections between each resonator, necessary in order to equalize ground potential across the two-dimensional lattice. Despite the lower kinetic inductance of NbTiN, the kinetic contribution still dominates the inductance. Because the kinetic inductance is intrinsically current-dependent, the resonances inherit a Kerr-type nonlinearity \cite{frascaHighkineticInductanceNbN2023}. All measurements presented here are performed at drive powers where the induced frequency shift remains well below the linewidth, i.e., in the linear regime.

Our CCA consists of 33 resonators arranged in an isotropic honeycomb lattice, terminated by a zigzag-type edge consisting of a single sublattice (Fig.~\ref{fig:CCA}c--e). The lattice therefore realizes an ``artificial graphene'' flake: as in natural graphene, the band structure consists of two bands touching at Dirac points \cite{castronetoElectronicPropertiesGraphene2009}, and a zigzag boundary hosts dispersionless (flat) edge modes existing in a limited area of the Brillouin zone \cite{fujitaPeculiarLocalizedState1996,dibenedettoEmergentCavityQEDDynamics2026}, which touch the bulk bands at the Dirac points. In our finite lattice, the bands are discretized, and the edge modes are expected to appear as states localized on the edge sublattice. To ensure that the edge resonators have the same resonant frequency as the bulk resonators, they are surrounded by ``ghost resonators''~\cite{jouannyHighKineticInductance2025}, resonator-shaped structures with a disconnected inductor that allow to preserve the same total capacitance for the edge resonators as for the bulk resonators. The two coupling ports follow the same ``ghost'' design. 
The fabricated device (Appendix~\ref{appendix:fabrication}) is wirebonded to a printed-circuit board (PCB) with two coplanar waveguides and inserted into a transmission RF setup in an attoDRY800 cryostat, which also allows optical access (Appendix~\ref{appendix:setup}). 

We present the RF transmission spectrum of our device in Fig.~\ref{fig:CCA}f.
All measurements shown here are performed at sample temperatures around $5$ to $6~\mathrm{K}$, leading to temperature-independent internal Q factors of $\approx 300$. For comparison, in a well-shielded dilution fridge setup at $20~\mathrm{mK}$, the internal Q factors of similar devices are above $10^4$. (Appendix~\ref{appendix:mk_measurement}). 
We show all additional device properties in Appendix~\ref{appendix:design}.

We model our system by constructing the Hamiltonian using standard circuit quantization:
\begin{equation}
\mathbf{H} / \hbar = \sqrt{\mathbf{C}^{-1}\mathbf{L}^{-1}},
\end{equation}
where $\mathbf{L}$ is a diagonal matrix containing inductances to ground, and $\mathbf{C}$ is a matrix containing the capacitances to ground and the coupling capacitances. These parameters are obtained from Ansys Maxwell Electrostatic simulations of the designed device. The circuit schematic of the lattice is shown in Fig.~\ref{fig:CCA}g. The resulting Hamiltonian can be represented as a tight-binding model:
\begin{equation}
\hat{H}
=
\sum_{j=1}^{N} \omega_j |j\rangle\langle j|
+
\sum_{j<k}^{N}
J_{j,k}
\left(
|j\rangle\langle k|
+\mathrm{h.c.}
\right),
\end{equation}
where $\omega_j$ is the on-site resonance frequency, $|j\rangle$ is the mode of resonator $j$, and $J_{j,k}$ is the coupling rate between sites $j$ and $k$. The diagonal $\omega_j$ terms are determined by the inductances to ground and the total capacitances, and the off-diagonal $J_{j,k}$ terms by the coupling capacitances and the inversion of the capacitance matrix. In practice, the matrix inversion, together with non-negligible parasitic capacitances, generates couplings beyond nearest neighbours, with $J'=0.13~J$ and $J''=0.045~J$ (Fig.~\ref{fig:CCA}h). We model dissipation using the non-Hermitian Hamiltonian:
\begin{equation}
\hat{H}^{\mathrm{NH}}
=
\hat{H} - i\frac{\kappa_{int}}{2}\mathbbm{1}-i\frac{1}{2}K^TK\,
\end{equation}
where $\kappa_{int}$ is the internal loss rate of the resonators, and $K$ is the coupling operator to the two external waveguide ports. From this object we can compute the scattering matrix components: 
\begin{equation}
    S(\omega) = iKG(\omega)K^T-\mathbbm{1}, \label{S_simul}
\end{equation}
where we use the frequency-domain Green's function:
\begin{equation}
    G(\omega)=\left [H^{\mathrm{NH}}-\omega\mathbbm{1}\right]^{-1}.
\end{equation}

The calculated spectrum obtained from Eq.~\ref{S_simul}, with homogeneous on-site resonance frequency $\omega_0=3.405~\mathrm{GHz}$ and nearest-neighbour coupling $J=83~\mathrm{MHz}$, is in good agreement with our measurement (see Fig.~\ref{fig:CCA}f) . We observe that about 14 modes couple to the ports (some of which are degenerate), while the remaining modes produce no visible features in transmission. We identify two main discrepancies between the data and the model prediction: we observe higher external coupling to higher-frequency modes, and the mode spacings are slightly different when compared to simulations. We attribute these differences to next-nearest coupling to the external ports, differences in the capacitive environment of edge and bulk resonators, and disorder, none of which are included in the model. The spectrum alone, however, provides no spatial information: it can neither verify that the realized lattice matches the design and supports the desired edge modes, nor quantify the disorder that may contribute to these discrepancies. We therefore image the eigenmodes, first with laser scanning microscopy and then through NV magnetometry.

\section{Mode Shape Measurement using Laser Scanning Microscopy}\label{sec:LSM}

We reconstruct the squared eigenmode amplitudes of the port-coupled modes using laser scanning microscopy \cite{zhuravelMeasurementLocalReactive2006,zhuravelUnconventionalRfPhotoresponse2012,wangModeStructureSuperconducting2019,morvanBulkPropertiesHoneycomb2022}. A weak focused laser ($1.2~\text{µW}$, $\approx 30~\text{µm}$ spot, $520~\text{nm}$ wavelength) locally raises the surface resistance of the NbTiN film, changing the transmission spectrum (Fig.~\ref{fig:LSM1}a). We model the effect of the laser by adding loss $\delta\kappa_j$ at the illuminated site $j$. In the single-mode approximation (drive near $\omega_u$, far from all other modes), the induced change in the scattering matrix is
\begin{equation}
    \delta S(\omega)=-\frac{\delta\kappa_j}{2}\frac{|u_j|^2}{(\omega_u-\omega-i\gamma_u^{tot})^2}\:K|u\rangle \langle u|K^T, \label{deltaS}
\end{equation}
where $|u\rangle$ is the eigenmode of the non-Hermitian Hamiltonian, $\gamma_u^{tot}$ is its total loss rate, and $K$ the port-coupling operator (Appendix~\ref{appendix:LSM}). The signal is thus proportional to the squared mode amplitude $|u_j|^2$, with a squared-Lorentzian lineshape that allows the modal contributions to be separated by peak integration. 
By scanning the laser across the lattice we measure the differential signal at each site (Fig.~\ref{fig:LSM1}b); across the 33 traces we identify 14 peaks, corresponding to the port-coupled modes. We construct the LSM response $|\psi|^2$ by integrating the differential signal over a frequency window of $16.5~\mathrm{MHz}$ around each peak. The resulting map is proportional to $|u_j|^2$ at each site; for degenerate modes, it is a weighted linear combination of the constituent eigenmodes.

Figures~\ref{fig:LSM1}c--h show the measured and simulated LSM spatial response at three distinct frequencies. We find good agreement between the two sets of spatial maps, validating our mode-shape measurement technique. Figures~\ref{fig:LSM1}c,e correspond to non-degenerate bulk eigenmodes, while Fig.~\ref{fig:LSM1}d corresponds to the degenerate subspace of the edge modes. In the non-degenerate case, we can use LSM to image single eigenmodes of the lattice, while in the degenerate case the profile is a linear combination of the constituent eigenmodes. We further illustrate this concept in Appendix~\ref{appendix:LSM_results}. Together, these results demonstrate that LSM images the eigenmodes of a compact superconducting resonator array with site-level resolution, providing direct spatial access to the mode structure of the lattice.

\begin{figure}[t]
    \includegraphics[width=0.95\linewidth]{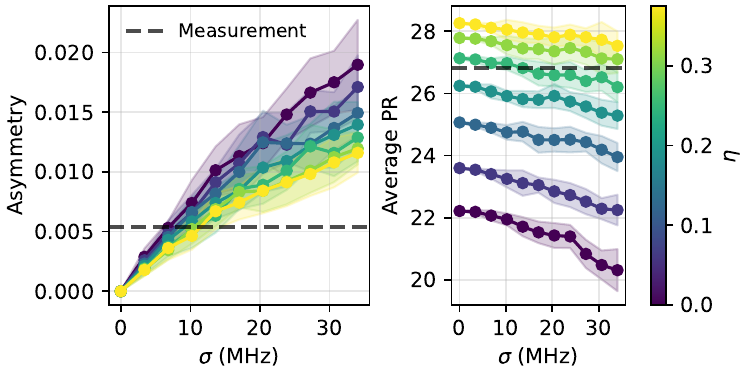}
    \caption{\textbf{Disorder estimation.} Asymmetry and average participation ratio (PR) of the eigenmodes as a function of on-site disorder. We average over multiple realizations of on-site disorder with standard deviation $\sigma$ and non-locality $\eta$, which is the strength of the perturbation on the nearest neighbours. 
    The measured values are shown with horizontal grey lines, and the simulated values with markers. The measured values are found to be consistent with a disorder of $\sigma\approx10~\mathrm{MHz}$ and a non-locality of $\eta\approx0.25$.}
    \label{fig:disorder}
\end{figure}

The mode profiles reveal details about disorder and localization in the lattice. Among the port-coupled modes we do not observe localized defect-like profiles, consistent with low disorder. We also observe symmetric mode profiles, as expected when the port coupling respects the mirror symmetry of the lattice. The primary difference between the measured and simulated profiles is in localization: the calculations predict more localized profiles. This is most apparent with the edge subspace, Fig.~\ref{fig:LSM1}d. In simulation, this profile is entirely localized on the sublattice making up the zigzag edge around the device; in measurements we observe partial sublattice polarization $\approx0.28$, which we calculate as
\begin{equation}
    P_{\mathcal{AB}} = \frac{\sum_{j\in \mathcal{A}} |\psi_{j}|^2/N_\mathcal{A}-\sum_{j\in \mathcal{B}} |\psi_{j}|^2/N_\mathcal{B}}{\sum_{j\in \mathcal{A}} |\psi_{j}|^2/N_\mathcal{A}+\sum_{j\in \mathcal{B}} |\psi_{j}|^2/N_\mathcal{B}},
\end{equation}
where $\psi_j$ is the reconstructed LSM amplitude on site $j$ (see Fig.~\ref{fig:LSM1}), and $\mathcal{A}$ and $\mathcal{B}$ denote the sublattices A and B, respectively. We attribute this effect to a non-local component in the LSM perturbation, as opposed to a deviation due to the realized lattice. 
In the simulation, we take this non-locality into account by introducing an additional perturbation on the nearest neighbours of the probed site, whose relative magnitude is denoted by $\eta$. Physically, this non-locality could arise because the SiO$_2$ layer on the Si substrate guides a fraction of the incident light to neighbouring sites, producing remotely induced losses.

\begin{figure*}[t]
    \includegraphics[width=0.99\linewidth]{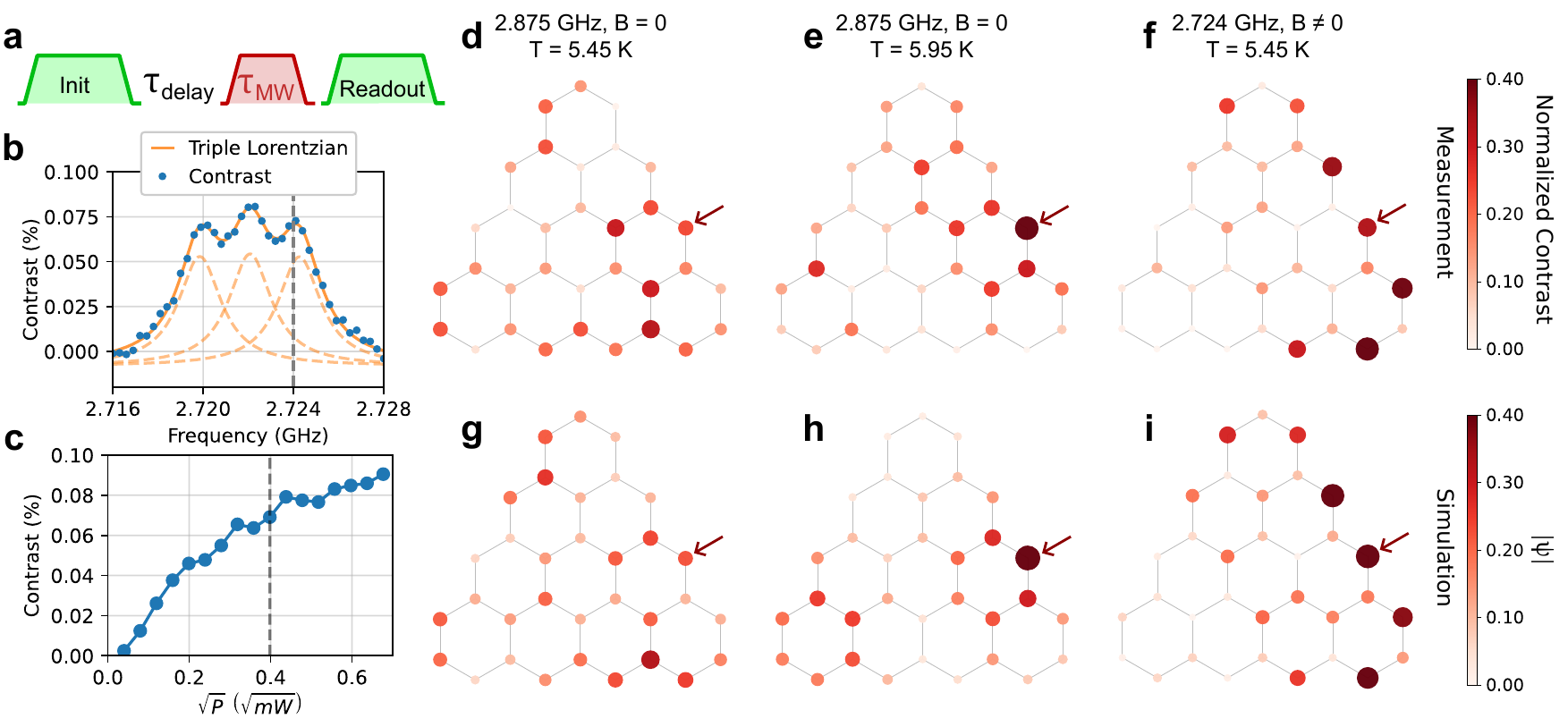}
    \caption{\textbf{NV-based field amplitude measurements.} \textbf{a} Laser and MW pulse scheme. \textbf{b} Pulsed optically detected magnetic resonance of a single NV population, measured on the input resonator, showing the hyperfine splitting due to the $^{14}\mathrm{N}$ nuclear spin. The hyperfine population at $2.724~\mathrm{GHz}$ was used for the measurements in panels \textbf{c} and \textbf{f}. \textbf{c} Photoluminescence (PL) contrast, measured on the input resonator, as a function of incident field amplitude for $\tau_{MW}=600~\mathrm{ns}$. We choose the input power highlighted by the vertical line for the measurement in panel \textbf{f}, corresponding to $-8~\mathrm{dBm}$ incident to the sample. \textbf{d--f} Spatial maps of the normalized PL contrast, measured under various driving conditions. The driven site is indicated by the arrows in the spatial maps. The maps are normalized such that $\sum_{j=1}^N|C_j|^2=1$. In panels \textbf{d, e} we use all overlapped NV populations at zero static magnetic field, and tune the spectral features of the device in resonance with the NVs by changing the base temperature. For \textbf{d} we use $\tau_{MW}=800~\mathrm{ns}$ and for \textbf{e} $\tau_{MW}=400~\mathrm{ns}$. In panel \textbf{f} we use a single hyperfine population tuned in resonance with the mode using a small static magnetic field. In all three maps we have $P_{\mathrm{in}}=-8~\mathrm{dBm}$. \textbf{g--i} Simulated normalized field amplitudes for the same driving conditions as the experiments.}
    \label{fig:NV}
\end{figure*}

Figure~\ref{fig:disorder} separates the respective impact of the on-site frequency disorder from the measurement non-locality: we compare the mode-averaged asymmetry and participation ratio of the measured modes with simulations, scanning the disorder standard deviation $\sigma$ and the non-locality $\eta$. We define the asymmetry $A$ and the participation ratio $\mathrm{PR}$ of a mode as
\begin{gather}
    A = \frac{\left | \sum_{j\in L} |\psi_{j}|^2-\sum_{j\in R} |\psi_{j}|^2 \right |}{\sum_{j} |\psi_{j}|^2} \quad\text{and}\\
    \mathrm{PR} = \frac{\left (\sum_{j} |\psi_{j}|^2 \right)^2}{\sum_{j} |\psi_{j}|^4},
\end{gather}
where $j\in L$~($j\in R$) denotes the sites to the left (right) of the lattice's axis of symmetry. In the simulated range, we observe that $A$ is more sensitive to on-site disorder, while $\mathrm{PR}$ responds mainly to the non-locality. Averaging both metrics over the port-coupled modes, we find $\sigma\approx10~\mathrm{MHz}$, corresponding to $\approx0.3\%$ of the bare center frequency, and $\eta\approx0.25$. This degree of non-locality reproduces a sublattice polarization of $\approx0.26$ for the simulated edge mode. The estimated disorder is comparable to that of NbN resonators on Si \cite{jouannyHighKineticInductance2025}. We have also verified that coupling disorder has smaller influence on both $A$ and $\mathrm{PR}$, thus it does not affect the above estimates (Appendix~\ref{appendix:coupling_disorder}). The separation between the roles of $\eta$ and $\sigma$, however, is an inference that relies on comparison with simulations of the lattice.

\section{Field Amplitude measurement using NV Magnetometry}\label{sec:NV}

We now map the local microwave magnetic-field amplitude of the driven lattice with NV magnetometry. We place a $\sim200~\text{µm}$ thick diamond slab containing a few-$\text{µm}$ thick layer of NV centers ($^{12}\mathrm{C}$-enriched from HiQuTe, 2--3 ppm NV concentration) \cite{balasubramanianUltralongSpinCoherence2009} directly on top of the superconducting lattice. The diamond's dielectric constant shifts the center frequency to $2.77~\mathrm{GHz}$ and the impedance to $Z=350~\Omega$. Importantly, a gradient in the diamond--resonator distance adds a linear gradient of on-site and coupling energies across the lattice, of $\approx 2.5\%$ in relative magnitude. This modifies the mode shapes, as confirmed by LSM measurements (Appendix~\ref{appendix:diamond_dielectric}). The NV spin ensemble and the device's modes can be tuned into resonance either by changing the base temperature (tuning the device resonant frequency) or a static magnetic field (tuning the NV spin transition). We optically address the spins using a custom microscope built around the cryostat (Appendix~\ref{appendix:setup}). The measurement relies on the pulsed sequence of Fig.~\ref{fig:NV}a, with the laser and the incident microwave drive pulsed in synchronization. The RF magnetic field generated around the inductive elements drives transitions within the spin triplet ground state of the NV centers \cite{dohertyNitrogenvacancyColourCentre2013}. We describe the spins using the Hamiltonian:
\begin{equation}
    {H}_{NV}= \hbar D S_z^2 + g_{NV} \mu_B \mathbf{S}\cdot \mathbf{B},
\end{equation}
with zero-field splitting $D/2\pi=2.877~\mathrm{GHz}$ \cite{dohertyNitrogenvacancyColourCentre2013}, $g_{NV}=2$, and $\mu_B/h=14~\mathrm{MHz/mT}$; $\mathbf{S}=(S_x, S_y, S_z)$ are the spin-1 Pauli operators, and $\mathbf{B}$ is the total magnetic field. Writing this field as $\mathbf{B}(t) = \mathbf{B}_0 + \mathbf{B}_{DUT}\cos(\omega t)$, with $\mathbf{B}_0$ the static tuning field and $\mathbf{B}_{DUT}$ the MW field produced by the probed resonator, a resonant drive coherently flips the spins at the Rabi frequency $\Omega_R=g_{NV}\mu_B B_{DUT,\bot}/\hbar$, i.e., proportional to the magnetic field component transverse to the NV quantization axis. Assuming complete initialization, the excited-state population after a pulse of duration $\tau_{MW}$ is $P_e = \frac{1}{2}-\frac{1}{2}\cos(\Omega_R\tau_{MW})$. Eventually, this results in a fluorescence contrast $C$ proportional to $P_e$~\cite{dohertyNitrogenvacancyColourCentre2013}. 

We experimentally implement the Rabi sequence from Fig.~\ref{fig:NV}a. The laser pulses used for spin initialization deposit approximately $5 \times 10^{-7}~\mathrm{J}$ into the device, generating a burst of nonequilibrium quasiparticles and a transient temperature rise in the superconducting structures \cite{liFastRecoveryNiobiumbased2025}. The nonequilibrium quasiparticles recombine within $\approx100~\mathrm{ns}$, while the return to full thermal equilibrium takes several milliseconds (Appendix~\ref{appendix:s21_recovery}); we therefore maintain $20~\mathrm{ms}$ dead time between laser pulses and delay the microwave drive by $19.9~\mathrm{ms}$, allowed by the long $T_1\approx 180~\mathrm{ms}$ of the NV ensemble (Appendix~\ref{appendix:T1}).

To validate our setup, we first perform pulsed optically detected magnetic resonance (ODMR), keeping a fixed $\tau_{MW}=600~\mathrm{ns}$. The results, shown in Fig.~\ref{fig:NV}b) for a non-zero $B_0$ and a single non-degenerate NV resonance, displays the expected hyperfine splitting due to the $^{14}\mathrm{N}$ nuclear spin. The low contrast is due to strong field inhomogeneities over the large laser spot: we only address spins that are physically close to the nanowire inductors (Appendix~\ref{appendix:field_inhom}). 

Our NV-based spatial mapping then relies on a simplified Rabi measurement: rather than recording full Rabi oscillations at each site, we perform pulsed ODMR with a fixed MW pulse length $\tau_{MW}$, comparing the fluorescence levels with the drive tuned on and off resonance; the resulting contrast reports the local field amplitude. We calibrate this pulse length by measuring the contrast $C_i$ at the input resonator site $i$ as a function of incident power $P$, see Fig.~\ref{fig:NV}c. Note that the field produced by the resonator is such that $B_{DUT} \propto \sqrt{P}$. The field inhomogeneity over the laser spot creates a broad distribution of Rabi frequencies, so in general the contrast is a monotonic but nonlinear proxy for the local field amplitude (the ideal, single spin model discussed above breaks down). 
As illustrated in Fig.~\ref{fig:NV}c, we set the pulse duration and incident power such that the pulse area $\Omega_R \tau_{MW}$ on the input resonator is sensitive to $B_{DUT}$. Our calibration also shows that the dependence can actually be considered as linear for field amplitudes smaller than in the input resonator.
Thus, keeping $\tau_{MW}$ and $P$ fixed, a single pulsed ODMR measurement per site yields a spatial map of the contrast, proportional to the local field amplitude under the chosen driving condition. Note that the chosen value for $\tau_{MW}$ also far exceeds the resonator ring-up time, such that our pulsed measurement probes the driven steady state of the array rather than its transient response. 


Figures~\ref{fig:NV}d--f show the normalized contrast maps under three driving conditions, alongside the simulated normalized steady-state field amplitudes (Fig.~\ref{fig:NV}g--i), which we compute from the Green's function as $a(\omega) = i G(\omega) K^T a_{in}(\omega)$, where $a_{in}(\omega)$ is the incident field vector, and $G(\omega)$ is constructed with the on-site and coupling energy gradient taken into account. In Figs.~\ref{fig:NV}d and e, we drive the lattice at zero static field and tune two different bulk modes into resonance with the NVs by changing the base temperature of the device. Fig.~\ref{fig:NV}d shows a bulk mode with complex structure, while Fig.~\ref{fig:NV}e shows a mode with a clear nodal line parallel to the lattice edge (and orthogonal to the diamond-resonator distance gradient). In both cases, the measured maps reproduce the simulated field profiles. 

In Fig.~\ref{fig:NV}f we tune the NVs with a static field, and drive the lattice nearly resonantly with an edge mode. Because the gradient in the diamond-resonator distance breaks the axial symmetry, the edge mode is no longer accessible with our transmission-type LSM measurement (see Fig.~\ref{fig:LSM2}), but can be resolved by NV magnetometry. The measured map is strongly localized on the edge sublattice; from the contrast map, we estimate the sublattice polarization to be $0.75$, close to the simulated $0.77$. The value is not corrected for the slightly nonlinear contrast response or the varying diamond--resonator distance (Appendix~\ref{appendix:diamond_dielectric}), which bias the estimate in opposite directions. The deviation from the maximal value of $\approx0.9$ is due to a detuning of $\approx 11~\mathrm{MHz}$ from the edge mode's frequency, which was present in the experiment due to the limited accuracy of tuning the bias magnetic field with a permanent magnet.

\section{Discussion and Conclusion}\label{sec:conclusion}


The two techniques studied here probe complementary observables of the array: LSM images the spatial structure of the eigenmodes, while NV magnetometry maps the local microwave magnetic-field amplitude of the driven lattice. 
On the one hand, LSM accesses many eigenmodes in a single measurement, but it is perturbative. Moreover, the optically-induced loss of superconductivity can extend well beyond the illuminated spot because of substrate-guided light. 
It is thus necessary to discriminate such measurement artifacts from structural features of the fabricated lattice.
We address this challenge by computing the asymmetry and participation ratio from the measured maps (Fig.~\ref{fig:disorder}) and comparing them with simulations. 
On the other hand, NV magnetometry captures the driven response directly in the unperturbed state (after allowing full recovery of the superconductivity), but at the cost of longer integration times and the need to tune the NV transition into resonance with the drive. 
The edge-mode sublattice polarization estimated from the NV contrast map ($0.75$, close to the simulated $0.77$) confirms that the reduced polarization observed with LSM ($\approx0.28$) originates primarily from the non-locality of the LSM probe rather than from the realized lattice. Using this spatially resolved information, we also establish an upper bound of $0.3\%$ on the on-site frequency disorder and find no evidence of defect modes. The agreement between the two methods validates the successful implementation of the target tight-binding Hamiltonian with the fabricated CCA, and establishes a combined characterization approach for compact superconducting resonator arrays.

While the present work focuses on the linear regime, the kinetic inductance of these resonators provides an intrinsic Kerr nonlinearity~\cite{frascaHighkineticInductanceNbN2023}, making them well-suited to the study of many-body driven-dissipative phenomena~\cite{beaulieuObservationFirstSecondorder2025, castillo-morenoExperimentalObservationMultimode2026}. In the nonlinear regime, the laser perturbation in LSM no longer translates into a change $\delta S$ proportional to the local mode occupation. Using NV magnetometry thus offers an interesting perspective for gaining spatially-resolved information on interacting photonic many-body states. Complete reconstruction of the local wavefunction can further be performed with phase-sensitive NV-based protocols, with excellent frequency resolution~\cite{staudenmaierPhasesensitiveQuantumSpectroscopy2021, kitamuraRobustGigahertzrangeAc2026}.
Looking ahead, increasing the coupling of large NV ensembles to the resonators would enable the realization of a hybrid spin--superconducting quantum simulator~\cite{zouImplementationDickeLattice2014, zhouSimulatingLipkinMeshkovGlickModel2017, zhuInterplayQuantumPhase2020}.


\section*{Acknowledgements}\label{sec:acknowledgements}

The project was supported by the Swiss National Science Foundation (SNSF) through Grant No. 216406. The authors gratefully acknowledge the Center of MicroNanoTechnology (CMi) at EPFL for providing cleanroom facilities and technical support.



\appendix

\section{Single resonator spectrum}\label{appendix:mk_measurement}

\begin{figure}[ht]
    \includegraphics[width=0.8\linewidth]{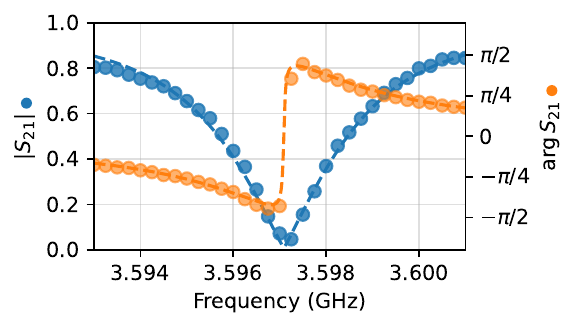}
    \caption{\textbf{Single resonator spectrum.} The $S_{21}$ magnitude and phase of a single resonator, side-coupled to a waveguide measured at $20~\mathrm{mK}$. The data (circles) are fitted with a complex circle fitting algorithm (dashed line) to extract the loaded, coupling, and internal quality factors.}
    \label{fig:mk_spectrum}
\end{figure}

We measure the spectrum of a single resonator side-coupled to a coplanar waveguide in a dilution fridge at $20~\mathrm{mK}$ (Fig.~\ref{fig:mk_spectrum}). We fit the data using a complex circle fitting algorithm \cite{probstEfficientRobustAnalysis2015}, and extract a loaded quality factor of $Q_L=747.2\pm1.9$, a coupling quality factor of $Q_c=756.5\pm1.4$, and an internal quality factor of $Q_i=5.9\times10^4\pm8.9\times10^3$. 

\section{Device design}\label{appendix:design}

Table~\ref{tab:CCAparams} summarizes the design parameters of the two-dimensional coupled cavity array presented in the main text.

\begin{table}[t]
\caption{\label{tab:CCAparams}Design parameters of the two-dimensional coupled cavity array.}
\begin{ruledtabular}
\begin{tabular}{@{}ll@{}}
Parameter & Value \\
\hline
\multicolumn{2}{l}{\textit{Lattice}} \\
Number of resonators & 33 \\
Lattice geometry & isotropic honeycomb \\
Edge termination & zigzag \\
Coupling ports & 2 \\
\multicolumn{2}{l}{\textit{NbTiN film}} \\
Critical temperature $T_c$ & $\approx 12~\mathrm{K}$ \\
Normal state sheet resistance & $\approx 63~\Omega$ \\
Kinetic inductance & $7~\mathrm{pH/\square}$ \\
\multicolumn{2}{l}{\textit{Unit cell geometry}} \\
Hexagonal cell side length & $100~\mathrm{\mu m}$ \\
Capacitive pad width & $18~\mathrm{\mu m}$ \\
Inductive nanowire width & $600~\mathrm{nm}$ \\
\multicolumn{2}{l}{\textit{Resonator (without diamond)}} \\
Inductance $L$ & $20~\mathrm{nH}$ \\
Kinetic to geometric inductance ratio & $\approx20$ \\
Total capacitance $C_{\Sigma}$ & $109~\mathrm{fF}$ \\
Resonance frequency $f_0$ & $3.405~\mathrm{GHz}$ \\
Impedance $Z$ & $427~\Omega$ \\
\multicolumn{2}{l}{\textit{Resonator (with diamond)}} \\
Total capacitance $C_{\Sigma}$ & $165~\mathrm{fF}$ \\
Resonance frequency $f_0$ & $2.77~\mathrm{GHz}$ \\
Impedance $Z$ & $350~\Omega$ \\
\multicolumn{2}{l}{\textit{Capacitance ratios}} \\
$C_c/C_g$ & $\approx 0.049$ \\
$C_c'/C_c$ & $\approx 0.096$ \\
$C_c''/C_c$ & $\approx 0.066$ \\
\multicolumn{2}{l}{\textit{Coupling rates}} \\
$J'/J$ & $0.13$ \\
$J''/J$ & $0.045$ \\
\multicolumn{2}{l}{\textit{Dissipation}} \\
Internal $Q$ ($5~\mathrm{K}$, unshielded) & $\approx 300$ \\
Internal $Q$ ($20~\mathrm{mK}$, shielded) & $\approx 10^4$ \\
Mode linewidth & $\approx 10~\mathrm{MHz}$ \\
Spin linewidth ($\mathrm{B}=0$) & $\approx 5~\mathrm{MHz}$ \\
Spin linewidth ($\mathrm{B}\neq0$) & $\approx 2~\mathrm{MHz}$ \\
\end{tabular}
\end{ruledtabular}
\end{table}

\section{Device fabrication}\label{appendix:fabrication}

The device is fabricated on a Si substrate covered with a 200 nm SiO$_2$ layer, coated with a 25 nm NbTiN film deposited by bias sputtering. The film has a critical temperature $T_c\approx12~\mathrm{K}$ and a kinetic inductance of $7~\mathrm{pH/\square}$ (Table~\ref{tab:CCAparams}). The full lattice is patterned in a single electron-beam lithography step using ZEP520 resist, and the pattern is transferred into the NbTiN film by reactive ion etching in an $\mathrm{SF_6}$ plasma. The chip is mounted on a copper sample holder alongside a printed circuit board carrying two coplanar waveguides, which are wirebonded to the two ports of the lattice.

\section{Derivation of the LSM response}\label{appendix:LSM}
Introducing local loss $\delta\kappa_j$ at site $j$ yields a rank-1 perturbation to the non-Hermitian Hamiltonian:
\begin{equation}
\delta H^{\mathrm{NH}} = -i\frac{\delta\kappa_j}{2}|j\rangle\langle j|,
\end{equation}
resulting in a correction to the Green's function:
\begin{multline}
    G'(\omega)=\left [H^{\mathrm{NH}}-i\frac{\delta\kappa_j}{2}|j\rangle\langle j|-\omega\mathbbm{1}\right]^{-1}=\\
=G(\omega)+i\frac{\delta\kappa_j}{2}\frac{G(\omega)|j\rangle\langle j|G(\omega)}{1-i\frac{\delta\kappa_j}{2}\langle j|G(\omega)|j\rangle},
\end{multline}
where we used the Sherman-Morrison formula. To first order in $\delta\kappa_j$, the correction to the scattering matrix is:
\begin{equation}
    \delta S(\omega)=-\frac{\delta\kappa_j}{2}K\:G(\omega)|j\rangle\langle j|G(\omega)K^T.
\end{equation}
We can expand $G(\omega)$ on the eigenmodes of the lattice:
\begin{equation}
    G(\omega) = \sum_{n}\frac{|n\rangle\langle n|}{\omega_n-\omega-i\gamma_n^{tot}},
\end{equation}
where $\omega_n-i\gamma_n^{tot}$ is the eigenvalue of mode $n$. In a single-mode approximation around eigenmode $|u\rangle$ (i.e. where $\Delta_u=\omega_u-\omega\approx0$ and all other $\Delta_n \gg 0$) we can simplify $G(\omega)$:
\begin{equation}
    G(\omega)=\frac{|u\rangle \langle u|}{\omega_u-\omega-i\gamma_u^{tot}}.
\end{equation}
The correction to the scattering-matrix is then:
\begin{equation}
    \delta S(\omega)=-\frac{\delta\kappa_j}{2}\frac{|u_j|^2}{(\omega_u-\omega-i\gamma_u^{tot})^2}\:K|u\rangle \langle u|K^T.
\end{equation}

\section{LSM and NV magnetometry measurement setup}\label{appendix:setup}

The measurement setup is based on an attoDRY800 cryostat with a base temperature of $\approx5~\mathrm{K}$, which provides optical access to the sample. Two RF lines (CryoCoax BCB019 steel coaxial cables) serve as input and output lines for the device, and are connected to a VNA (R\&S ZNL6) for transmission measurements or to a vector MW source (R\&S SMBV100B) for the NV measurements. The lines are attenuated by $20~\mathrm{dB}$ at the cold stage to reduce thermal noise. To enable spatial scanning we use a piezo stage, which necessitates the use of flexible cryogenic cables (el-spec MK5005) for the final coaxial connections.

For the optical experiments we use a custom confocal microscope. The laser (Cobolt 06-MLD, $520~\mathrm{nm}$) is coupled into a single-mode fiber, and the beam is routed through alignment mirrors, a filter wheel for attenuation, and a dichroic mirror. The beam is focused by a lens outside the cryostat window to a spot size of $\approx30~\text{µm}$. In the NV magnetometry experiment, the emitted PL is collected by the same lens, then separated from the excitation beam by the dichroic mirror, and filtered by a longpass filter. The filtered light is then coupled into a multimode fiber, and detected by an avalanche photodiode (Thorlabs APD410A). The NV experiment utilizes a pulsed measurement scheme where both the laser and incident microwave drive are pulsed in synchronization. The experiment is controlled by an FPGA-based multi-instrument device (Moku Pro), which generates the pulses controlling the laser and the microwave source, and performs boxcar averaging of the fluorescence (where we integrate the first $90~\text{µs}$ of each pulse).

\section{Comparison of the LSM spatial response to the lattice eigenmodes}\label{appendix:LSM_results}

\begin{figure*}[t]
    \includegraphics[width=1.0\linewidth]{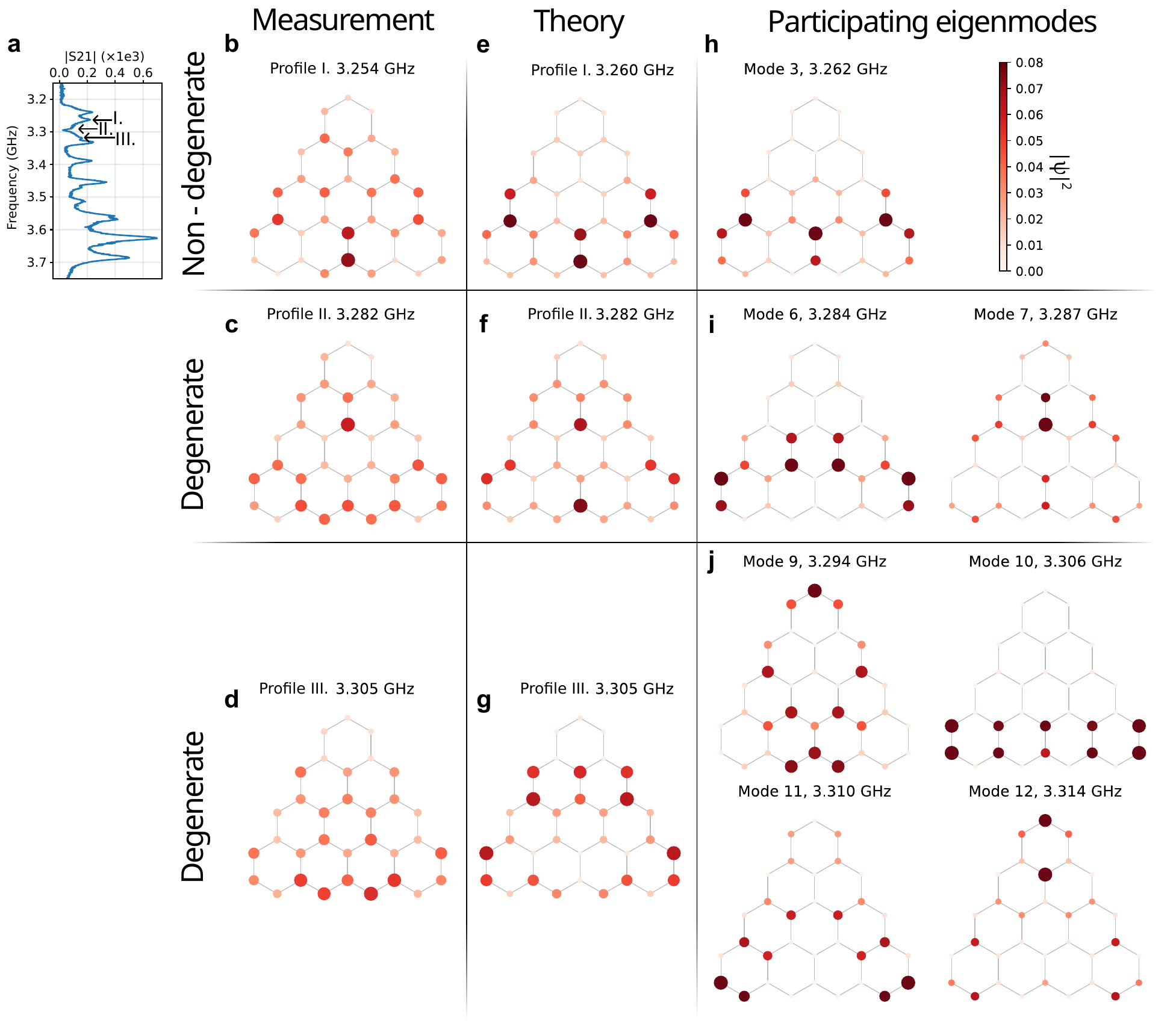}
    \caption{\textbf{Comparison of LSM profiles to lattice eigenmodes.} \textbf{a} The RF spectrum of the device with a non-degenerate (I.), and two degenerate (II. and III.) modes labeled. \textbf{b--d} The measured LSM profiles of the three modes. \textbf{e--g} The corresponding simulated LSM profiles. \textbf{h--j} Constituent lattice eigenmodes of the three LSM profiles. In a non-degenerate case, the LSM profiles strongly match a single eigenmode, while in a degenerate case the profile is a combination of several eigenmodes. In most cases, we can easily identify features of the constituent eigenmodes in the measured LSM profiles (center row), while in a multiple-degenerate case (bottom row) it is difficult to identify the constituent modes.}
    \label{fig:LSM_eigenmodes}
\end{figure*}

We present a comparison of selected LSM profiles to the corresponding lattice eigenmodes in Fig.~\ref{fig:LSM_eigenmodes}.

\begin{figure*}[t]
    \includegraphics[width=0.98\linewidth]{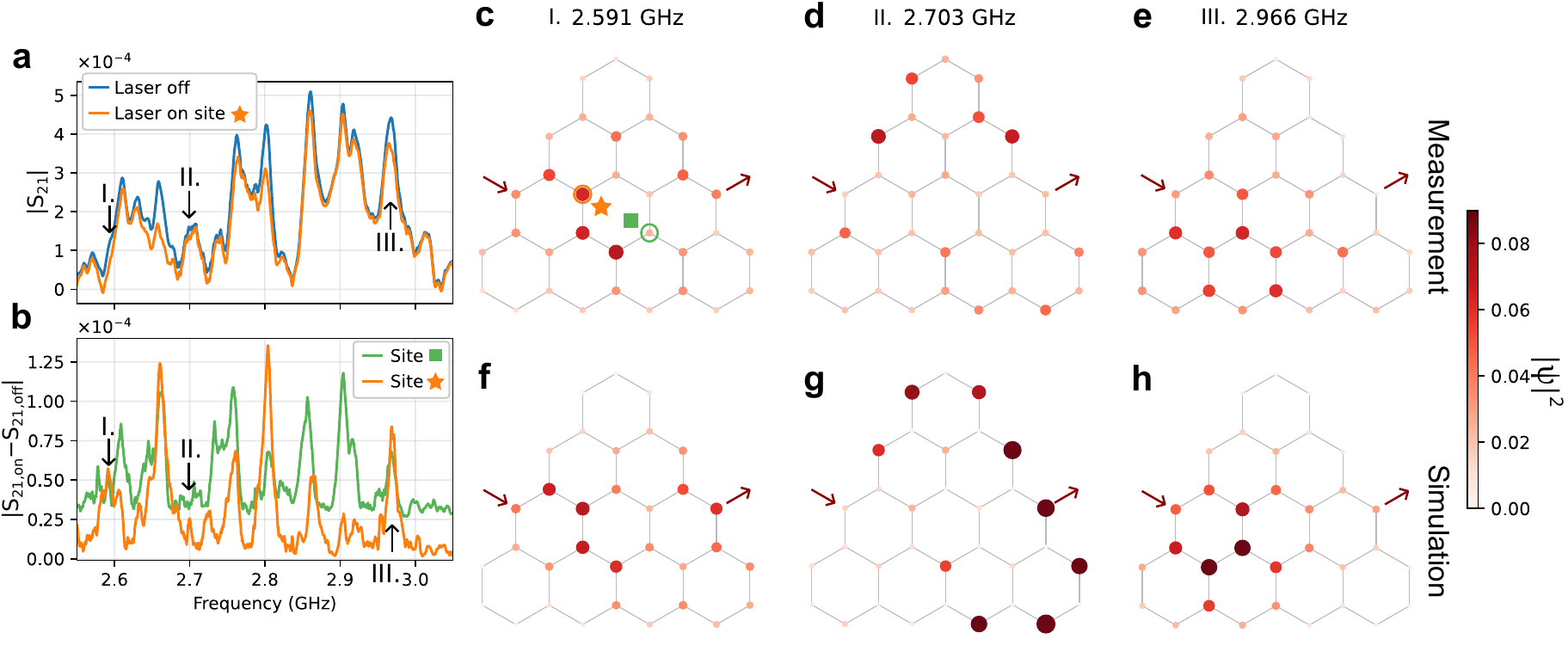}
    \caption{\textbf{LSM measurement of the device with diamond on top.} \textbf{a} The RF spectrum of the device without, and with a CW $1.2~\text{µW}$ laser illuminating the specified site. \textbf{b} LSM differential signal for two specified sites. For each site the peaks in the differential signal are identified and integrated, resulting in a spatial map. \textbf{c-e} LSM spatial maps of three modes. The modes are highlighted with arrows in panel \textbf{b}. \textbf{f-h} Simulated LSM profiles of the shown modes.}
    \label{fig:LSM2}
\end{figure*}

\begin{figure}[h]
    \includegraphics[width=0.95\linewidth]{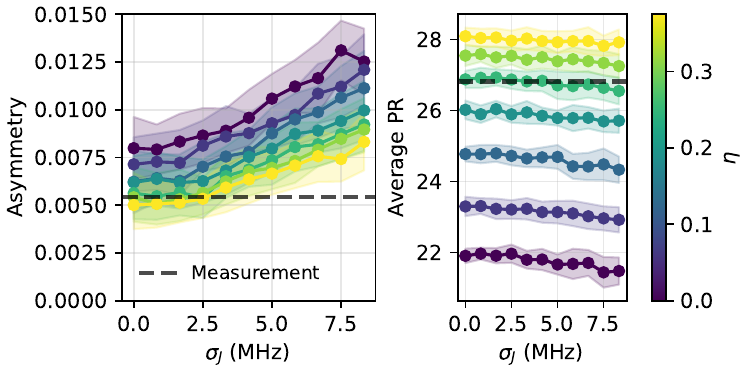}
    \caption{\textbf{Coupling disorder estimation.} Asymmetry and average participation ratio (PR) of the LSM spatial response as a function of coupling disorder, with 0.3\% relative on-site disorder present. We average over multiple realizations of coupling disorder with standard deviation $\sigma_J$ and non-locality $\eta$, which is the strength of the perturbation on the nearest neighbours. 
    The measured values are shown with horizontal grey lines, and the simulated values with markers. The measured values are found to be consistent with a coupling disorder of $\sigma_J\lessapprox2.5~\mathrm{MHz}$ and a non-locality of $\eta\approx0.25$.}
    \label{fig:coupling_disorder}
\end{figure}

\section{Estimation of coupling disorder}\label{appendix:coupling_disorder}

We analyse the coupling disorder $\sigma_J$ in Fig.~\ref{fig:coupling_disorder}, assuming the presence of 0.3\% relative on-site frequency disorder. We observe that below $\sigma_J=2.5~\mathrm{MHz}$ the asymmetry plateaus, meaning we cannot distinguish between these values of disorder. This allows us to place an upper bound of $\sigma_J\lessapprox2.5~\mathrm{MHz}$ on the coupling disorder.

\section{Altered properties due to diamond placement}\label{appendix:diamond_dielectric}

Diamond's dielectric constant is $\epsilon_r=5.7$, which drastically changes the capacitive environment compared to vacuum. The N-grease we use to fix the diamond has $\epsilon_r=2.3$. Inhomogeneity in the device-diamond distance thus causes a gradient in dielectric constant experienced by the resonators. In our device the distance varies from $\approx 250~\mathrm{nm}$ to $1~\text{µm}$ (which we measure using white light interferometry), resulting in a $\approx2.5\%$ gradient in onsite energies and coupling rates.
Figure~\ref{fig:LSM2} shows the LSM results obtained with the diamond on top of the device. The on-site energy and coupling gradient shifts the device center frequency and modifies the mode shapes, which we characterize using LSM. We find that the measured profiles are in agreement with the simulated ones, except for an edge mode (Fig.~\ref{fig:LSM2}d and g), where due to the on-site energy gradient the mode is localized on the right side of the lattice and does not couple to the left-side port, and is therefore not visible in the LSM measurement. In Fig.~\ref{fig:NV}f we drive the device near-resonantly with this mode and reconstruct the field amplitudes using NV magnetometry, obtaining good agreement with simulations.

The diamond distance gradient could be mitigated by optimizing the diamond placement method, e.g. by employing metal-metal thermocompression bonding \cite{riedelScalablePhotonicQuantum2025}. While the method is non-perturbative during the interrogation interval, the inclusion of diamond must be taken into account during the design phase.

\section{Field inhomogeneity and Rabi frequency distribution}\label{appendix:field_inhom}

The inductive element consists of a meander nanowire. Due to the adjacent antiparallel wires the field is cancelled in the far-field, therefore the NV centers must be positioned as close as possible to the resonators. The strong inhomogeneities in magnetic field lead to a wide distribution of Rabi frequencies. In our NV-based field amplitude measurement we detect fluorescence contrast from NVs that have the highest Rabi frequency (as we operate at or below $\pi/2$), but at longer pulse durations we see a monotonous rise in contrast as opposed to saturation.

\section{Effects of strong laser perturbation}\label{appendix:s21_recovery}

\begin{figure}[ht]
    \includegraphics[width=0.98\linewidth]{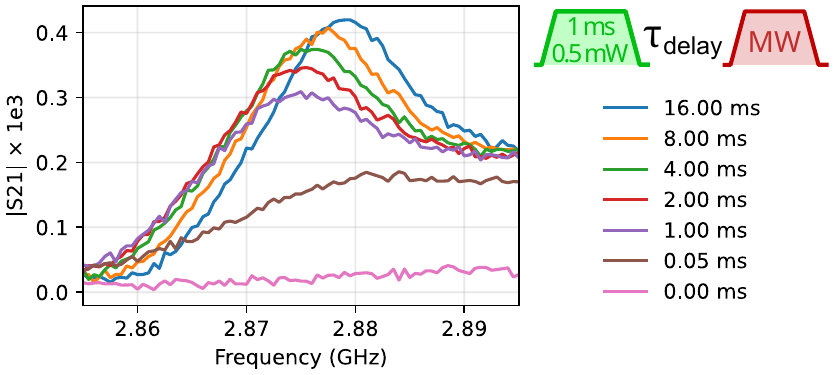}
    \caption{\textbf{Time-resolved RF spectroscopy.} $|S_{21}|$ of a single mode near $2.88~\mathrm{GHz}$ as a function of frequency, measured at different delays $\tau_{\mathrm{delay}}$ after a strong laser pulse ($0.5~\mathrm{mW}$, $1~\mathrm{ms}$, inset) applied to the resonator adjacent to the input port. The resonance is suppressed immediately after the pulse and recovers within $\approx10~\mathrm{ms}$. To measure the recovery, we use the continuous wave mode of the VNA at an IF bandwidth of $10~\mathrm{kHz}$, repeated for each frequency point.}
    \label{fig:s21_recovery}
\end{figure}

In the NV measurement sequence, the laser pulses used for spin initialization deposit approximately $5 \times 10^{-7}~\mathrm{J}$ into the device (Sec.~\ref{sec:NV}). To characterize the resulting perturbation, we apply a strong ($0.5~\mathrm{mW}$, $1~\mathrm{ms}$) laser pulse to the resonator adjacent to the input port and monitor the transmission spectrum as a function of the delay $\tau_{\mathrm{delay}}$ after the pulse, shown in Fig.~\ref{fig:s21_recovery}. Immediately after the pulse ($\tau_{\mathrm{delay}}=0$), the mode transmission is almost completely suppressed: the added quasiparticles and the transient temperature rise increase the surface resistance, washing out the mode. The transmission partially recovers within the first $50~\text{µs}$, consistent with the fast recombination of the nonequilibrium quasiparticles, while the full recovery of the resonance frequency and amplitude takes $\approx10~\mathrm{ms}$, reflecting the slower relaxation of the lattice temperature back to the base temperature. These time scales justify the $20~\mathrm{ms}$ dead time between laser pulses used in the NV measurement sequence (Sec.~\ref{sec:NV}), which ensures that the device has recovered before the microwave drive is applied.

\begin{figure}[ht]
    \includegraphics[width=0.8\linewidth]{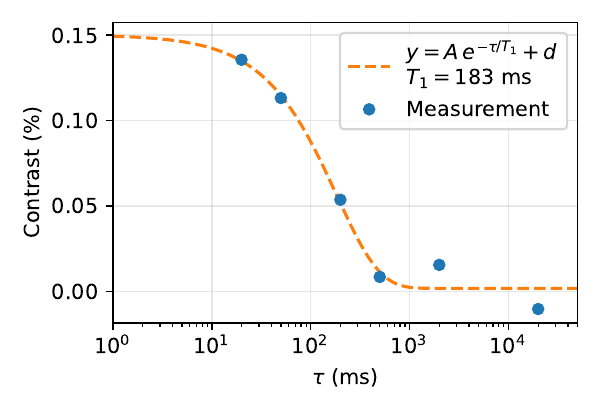}
    \caption{\textbf{$T_1$ measurement.} Fluorescence contrast of the NV ensemble as a function of delay after a $\pi$-pulse, measured on an identical diamond sample under conditions similar to the main experiment. The dashed line is an exponential fit yielding $T_1\approx 183~\mathrm{ms}$.}
    \label{fig:T1}
\end{figure}

\section{\texorpdfstring{$T_1$ measurement}{T1 measurement}}\label{appendix:T1}

We measure the $T_1$ of the NV ensemble of an identical diamond sample ($^{12}\mathrm{C}$-enriched from HiQuTe, 2-3 ppm NV concentration) under similar conditions as the main experiment. The results are shown in Fig.~\ref{fig:T1}. The measurement is performed on a single hyperfine population in the presence of a weak static magnetic field. We use a pulsed measurement scheme where we first initialize the spins to the ground state using a 2~mW, $1~\mathrm{ms}$  laser pulse, then apply a $\pi$-pulse to transfer the population to the excited state, and finally measure the population after a variable delay using another $1~\mathrm{ms}$ laser pulse. The contrast is extracted by comparing pulse trains with resonant and non-resonant MW. The fluorescence contrast is recorded as a function of the delay time, and we fit the resulting decay curve to an exponential function to extract the $T_1$ time. We find $T_1\approx 183~\mathrm{ms}$, which is consistent with previous measurements of NV centers in similar diamond samples at low temperatures \cite{jarmolaTemperatureMagneticFieldDependentLongitudinal2012}.


\bibliography{refs}



\end{document}